\documentclass[twocolumn,showpacs,amsmath,amssymb,pra,superscriptaddress]{revtex4}

\usepackage{graphicx}
\usepackage{dcolumn}
\usepackage{bm}
\usepackage{subfigure}

\begin{document}

\title{Peculiar chemical bonding between thorium and a carbon hexagon\\
in carbon nanomaterials}

\author{A. V. Bibikov}

\affiliation{Skobeltsyn Institute of Nuclear Physics Lomonosov
Moscow State University, 119991 Moscow, Russia}

\affiliation{National Research Nuclear University MEPhI, 115409,
Kashirskoe shosse 31, Moscow, Russia}

\author{A. V. Nikolaev}

\affiliation{Skobeltsyn Institute of Nuclear Physics Lomonosov
Moscow State University, 119991 Moscow, Russia}

\affiliation{National Research Nuclear University MEPhI, 115409,
Kashirskoe shosse 31, Moscow, Russia}

\affiliation{School of Electronics, Photonics and Molecular Physics, Moscow Institute of Physics and Technology, 141700, Dolgoprudny, Moscow region, Russia}

\author{E.~V.~Tkalya}
\email{tkalya_e@lebedev.ru}

\affiliation{P.N. Lebedev Physical Institute of the Russian
Academy of Sciences, 119991, 53 Leninskiy pr., Moscow, Russia}

\affiliation{National Research Nuclear University MEPhI, 115409,
Kashirskoe shosse 31, Moscow, Russia}

\affiliation{Nuclear Safety Institute of RAS, Bol'shaya Tulskaya
52, Moscow 115191, Russia}


\begin{abstract}
We explore an unusual nature of chemical bonding of the thorium
atom with a ring of six carbon atoms (hexagon) in novel carbon materials.
Our {\it ab initio} calculations of Th-based metallofullerenes
(Th@C$_{60}$, Th@C$_{20}$) and Th bound to benzene or coronene at
the Hartree-Fock level with the second order perturbation (MP2)
correction accounting for the van der Waals interactions,
demonstrate that the optimal position of the thorium atom is where
it faces the center of a hexagon and is located at a distance of
2.01--2.07~{\AA} from the center. For Th encapsulated in C$_{60}$ it is found at 2.01~{\AA},
whereas the other local energy minima are shifted
to larger energies (0.22~eV and higher). Inside C$_{60}$ the highest local minimum
at 1.17~eV is observed when Th faces the center of the five member carbon ring
(pentagon). Based on our calculations for Th with benzene and coronene
where the global minimum for Th corresponds to its position at 2.05~{\AA} (benzene)
or 2.02~{\AA} (coronene) from the hexagon center, we conclude that a well
pronounced minimum is likely to present in graphene
and in a single wall carbon nanotube. 
The ground state of Th is singlet, other high spin states (triplet and quintet) 
lie higher in energy ($> 1.62$~eV).
We discuss a potential use
of the carbon nanomaterials with the $^{229}$Th isotope having the
nuclear transition of the optical range, for metrological
purposes.
\end{abstract}


\maketitle

\section{Introduction}
\label{sec:int}

Metallofullerenes --- endohedral molecular complexes with a few
metal atoms located inside the molecular cage --- can be
considered as building blocks for novel nano-materials with
promising properties, and as such they are the subject of
intensive scientific research \cite{Klaus}. Nowadays, the
metallofullerene family $M$@C$_n$ includes dozens of objects ($n=60$, 70, 84 etc.) with
more than 30 various metals ($M=$Li, Sc, Y, La etc.) \cite{Lu}. Endohedral
metallofullerenes (EMF) and their properties are an increasingly
important topic in physics, chemistry, material science and
biology. Having high electron mobility and low reorganization
energy, EMFs are intensively studied for example in search for
novel donor-acceptor dyads for new metallofullerene-based solar
cells \cite{Rudolf}. On the other hand, owing to their high proton
relaxivity lanthanide metallofullerenes are scrutinized for novel
forms for magnetic resonance imaging (MRI) \cite{Mikawa}. This can
lead to the application of EMFs as MRI contrast agents in biology.
Another application in biology involves the usage of EMFs for
radiotraces \cite{Cagle}. For example, radioactive $^7$Be@C$_{60}$
can be detected using radiochemical and radiochromatographic
techniques \cite{Ohtsuki}.

In this work as the encapsulated object we consider the atom of
thorium which belongs to the actinide series. For the first time light actinide
metallofullerenes were synthesized and isolated
with high-performance liquid chromatography in 2001, Refs.\ \cite{Akiyama,Akiyama-2}.
There, it has been shown that similarly to lanthanides, the encapsulated
actinide atoms are in the trivalent state. Thorium however differs
from the light actinides \cite{Akiyama-2}. Its oxidation state in
Th@C$_{84}$ is $4+$, implying a strong hybridization with the
electron cloud of the fullerene cage.
Very recently, in a series of experimental studies
mono-EMFs based on uranium and thorium have been successfully
synthesized and characterized by single crystal x-ray diffraction,
various spectroscopies, electrochemistry and numerical
calculations based on the density functional theory (DFT) \cite{Wang17,Jin19,Wang19}.
The DFT investigation of the interaction of Th with carbon cage structures
was continued in theoretical works \cite{Kamin,Jin17,Li-b,Li-a}.
Th-based EMFs in these studies include the Th@C$_{2n}$ family
with $2n=82$ \cite{Wang17}, 86 \cite{Wang19}, 76 \cite{Jin17,Jin19}, 74 \cite{Li-b}, 64-88 \cite{Li-a}, and others.

Among various applications, the choice of the thorium
atom is related to the possible usage of Th-based
metallofullerenes for metrology. The fullerene cage protects the
encapsulated atom from the chemical environment and reduces the
external interactions, which is beneficial for these purposes.
The $^{229}$Th isotope has a low energy nuclear transition in the
optical range \cite{Seif}, which can be used for a new laser
working on it \cite{Tkalya}, establishing a new (nuclear) standard
of chronography \cite{Peik}. For the solution of this practical
task one has to find an effective media for the $^{229}$Th
isotope, facilitating the nuclear transition. As possible
candidates we consider here mono-EMFs Th@C$_{60}$ and Th@C$_{20}$,
and explore theoretically their main electron properties.

Mono-EMFs are generally considered particularly suitable for elucidating metal-cage interactions.
The nature of the metal-cage interaction in EMFs is not completely understood: although
the bonding is believed to be mainly ionic, in many cases a substantial degree of covalency
has been observed with the valence electrons shared between the encapsulated atom and the carbon cage \cite{popov}.
For the thorium metallofullerenes Th@C$_{2n}$ studied recently by means of DFT theory
the Th-cage bonding is found to be polar covalent, with
an appreciable back-donation from cage (C$_{84}^{4-}$) to Th (Th$^{4+}$).
In many cases the Th atom is located at a center of a
hexagon, with shortest Th-C contacts at approximately 2.4-2.5~{\AA} \cite{Kamin,Jin19,Jin17,Wang19}.
In Th@$C_{3v}$(8)-C$_{82}$ however it has been concluded that Th resides
at the conjunction of three hexagons \cite{Wang17}.

In the present paper we investigate the nature of the Th-cage bonding in other fullerenes (Th@C$_{60}$ and Th@C$_{20}$)
at the Hartree-Fock level taking into account the weak dispersion forces by including the second order M{\o}ller-Plesset
perturbation theory (MP2). 
So far, the Th dispersion forces have been included in calculations in the form of
the Grimme correction only in Ref.\ \cite{Wang19}. 
In particular, we closely examine the character of the interaction between the Th atom and the six member carbon ring (hexagon)
which is the essential part of the bonding mechanism.
For that we consider the bond formation in fullerenes (C$_{60}$ and C$_{20}$), but also
in simpler molecules like benzene and coronene (plain and curved).
This investigation is helpful in transferring our results and conclusions with respect to thorium-cage bonding mechanism to other carbon materials like graphene and carbon nanotubes.
Earlier, chemical bonding of actinides (Th--Cm) with graphene-like molecules have been carried out by DFT method in Ref.\ \cite{Du17}.
However, in Ref.\ \cite{Du17} the thorium atom has been taken in magnetic -- triplet ($S=1$) or quintet ($S=2$) -- ground state whereas our analysis for Th given
in this paper clearly indicate the singlet ($S=0$) ground state for it.
Similar studies but with light alkali and alkali earth metal atoms have been carried out in \cite{Bodrenko12}
for determination of the effectiveness of the binding of molecular hydrogen.
It is worth noting that the binding of a metal atom on the graphene sheet is a subject of intense research
(see for example Ref.\ \cite{Liu} and references therein) with a potential application in carbon-based electronic devices.
The problem is also of fundamental interest from the viewpoint of adatom-graphene chemical bonding mechanism \cite{Dima}.

\section{Method}
\label{sec:model}

The calculations have been performed within {\it ab initio} Hartree-Fock (HF) method \cite{SO} as
implemented in Refs.\ \cite{Artemyev-05} and \cite{GAMESS}.
The adopted molecular basis sets were 6-31G$^{**}$ for carbon, and all-electron
augmented triple zeta valence quality basis set with polarization functions
(jorge-ATZP in the Basis Set Exchange site \cite{BSE})
for thorium \cite{JorgeT}. Note that unlike basis sets with
the pseudopotential treatment of the core states,
the jorge-ATZP allows for a full core polarization effect of the thorium atom.
The auxiliary resolution of identity basis sets were 0.003.RI for C, and aug-cc-pVTZ for Th.
For some precise calculations we have used a larger basis set -- aug-cc-pVTZ \cite{Dunn} --
also for carbon and hydrogen.
To account correlations responsible for the weak dispersion forces,
we have used the second order M{\o}ller-Plesset perturbation correction (MP2) \cite{SO}.

Thorium atom is generally non-magnetic and experimentally it has never been found in high spin states in molecular structures
or in solids. Our calculations 
with the singlet ($S=0$), triplet ($S=1$) and quintet ($S=2$) spin-adapted configurations for Th with the benzene have been fully confirmed
this expectation. The energy differences are 1.623 eV (triplet-singlet) and 2.025 eV (quintet-singlet) for the optimal geometries and
the jorge-ADZP basis set \cite{BSE}. Therefore, in the following we will use only the singlet spin-adapted configurations.

\section{Results and discussion}
\label{sec:model}

\subsection{Th@C$_{60}$: thorium inside the C$_{60}$ molecular cage}
\label{ThC60}

\begin{figure}
\resizebox{0.45\textwidth}{!} {
\includegraphics{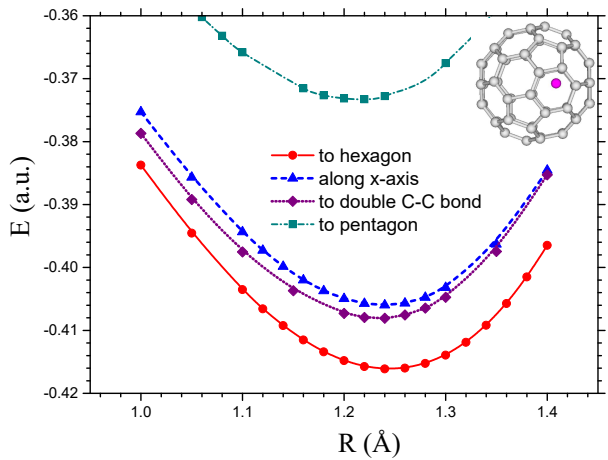}
}
\caption{
Binding energy $E_b$ of the Th@C$_{60}$ molecular complex versus the Th displacement $R$ from the fullerene center
in various directions:
1) red line (circles) -- to the center of a hexagon,
2) purple line (diamonds) -- to the middle point of the double C-C bond (fusing two hexagons of C$_{60}$),
3) blue line (triangles) -- along the $x$-axis,
4) dark cyan line (squares) -- along the $z$-axis (to the center of a pentagon).
} \label{fig1}
\end{figure}

We consider the C$_{60}$ molecule in the orientation where its highest 5-fold rotation axis lies along the $z$ Cartesian
direction and goes through the center of a pentagon. The $y$ axis goes through the middle point of the double C-C bond fusing two hexagons,
and the $x$-axis being perpendicular to the $y$ and $z$ axes, passes through a hexagon.
Then we locate the thorium atom at the center of C$_{60}$ and displace it in various directions to find its preferred energy position
inside the fullerene. For that purpose we compute the binding energy $E_b$ defined according to
\begin{eqnarray}
    E_{b} = E(\textrm{Th@C}_{60}) - E(\textrm{Th}) - E(\textrm{C}_{60}) . \nonumber
\end{eqnarray}
The corresponding energy plots for the Th@C$_{60}$ complex are shown in Fig.~\ref{fig1},
while the main results are given in Table \ref{tab1}.
%
%

We have found that the energy is lowered when the Th atom is displaced
from the molecular center by a distance of $1.211-1.244$~{\AA}.
The situation is very different from the fullerene complexes BeC$_{60}$ \cite{Bib1} and BeC$_{70}$ \cite{Bib2}
where the beryllium atom remains at the molecular center.
This is because beryllium has a small atomic radius of 1.05 {\AA} and its highest occupied molecular orbital (HOMO)
at -9.3~eV lies below the HOMO shell of C$_{60}$ (-7.74~eV) and C$_{70}$ (-7.14~eV), forbidding a charge transfer to
the fullerene and efficient chemical bonding.
Thorium atom with an effective atomic radius of 1.8~{\AA} and the $6d$ and $7s$ electron levels
lying higher in energy, interacts more strongly with the electron shell
of the C$_{60}$ fullerene.
Note that the energy minima shown in Fig.\ \ref{fig1} imply that the Th atom is located
at distances $2.31-2.34$~{\AA} from the C$_{60}$ fullerene sphere
passing through all carbon nuclei.

The lowest minimum corresponds to the optimal position of Th on the line
to the hexagon at $d_0^{min} = 2.01$~{\AA} from its center.
The geometry optimization of Th@C$_{60}$ confirms this position of the Th atom implying the $C_{3v}$
symmetry of the molecular system, Table \ref{tab1}.
The relatively large value of the Th-C bonds $l_{b}=$2.46 {\AA} here is a result of sharing
the chemical bonding of Th with its six nearest carbon neighbors.
%
\begin{table}
\caption{Thorium bonding with various carbon structures, MP2 calculations.
$d_0$ is the distance to the carbon ring center ({\AA}),
$d$(Th-C) is the closest thorium-carbon bond length ({\AA}), MO energies are in eV.
\label{tab1} }

\begin{ruledtabular}
\begin{tabular}{c | c  c  c }

                & Th@C$_{60}$  & ThC$_6$H$_6$   & ThC$_{24}$H$_{12}$   \\
\tableline
   symmetry     & C$_{3v}$       & C$_{6v}$       &  C$_{6v}$ \\
    $d_0$       & 2.006          & 2.050          &  2.018 \\
    $d$(Th-C)   & 2.459          & 2.504          &  2.470 \\
    HOMO $-$ 1  & -7.926 ($E$)   & -9.546 ($E_1$) & -6.834 ($E_1$) \\
     HOMO       & -5.220 ($E$)   & -4.529 ($E_2$) & -3.568 ($E_2$) \\
     LUMO       & -1.100 ($A_1$) & -0.045 ($A_1$) &  0.050 ($A_1$) \\
    LUMO $+$ 1  & -0.090 ($A_1$) &  0.485 ($A_1$) &  0.507 ($A_1$) \\
\end{tabular}
\end{ruledtabular}
\end{table}
Under the group $C_{3v}$ the three-fold degenerate $t_{1u}$ lowest unoccupied molecular orbital (LUMO) of C$_{60}$
splits into the $E + A_1$ levels of $C_{3v}$, with the two-fold degenerate $E$ state lying
lower in energy. This $E$ level of Th@C$_{60}$ is then completely occupied by the four electrons of thorium, Table \ref{tab1}.
The minima in other directions lie higher in energy, the closest local minimum (the purple line in Fig.\ \ref{fig1}) being at 0.22~eV.
The highest value of 1.17~eV corresponds to
the Th position at 2.10~{\AA} off the center of the pentagon, the dark cyan line in Fig.\ \ref{fig1}.

The lowest energy minimum of Th@C$_{60}$ is in good correspondence with the results of Kaminsk\'{y} et al. \cite{Kamin}
for the Th@C$_{84}$ fullerene with Th located at a center of hexagon with shortest Th-C contacts of 2.5~{\AA}.
The discrepancy with our bond length $l_b$ is only 1.6{\%}
although technically our calculations are different:
in Ref.\ \cite{Kamin} the PBE0 hybrid functional was used whereas in our case -- HF with the MP2 correction.
This minimum has also been reported for other DFT calculations in Ref.\ \cite{Wang19} for Th@C$_{86}$
(bond lengths 2.49-2.53 {\AA}),
in Ref.\ \cite{Jin19} for Th@C$_{76}$ (2.39-2.47 {\AA}).
It is therefore intrinsic to the interaction of Th with carbon cages of fullerenes.
In Th@C$_{82}$ however it has been found that Th is located at 2.34 {\AA} from a carbon atom
at the conjunction of three hexagons \cite{Wang17}.

\subsection{Th@C$_{20}$}
\label{ThC20}

\begin{figure}
\resizebox{0.43\textwidth}{!} {
\includegraphics{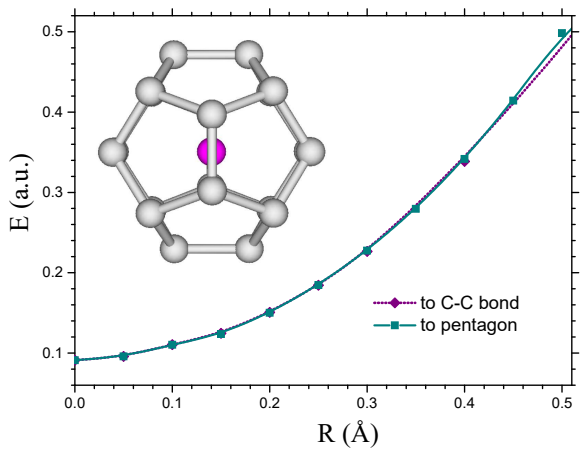}
}
\caption{
Binding energy $E_b$ of the Th@C$_{20}$ molecular complex versus the Th displacement $R$ from the fullerene center
in various directions:
1) dark cyan line (squares) -- to the center of a pentagon,
2) purple line (diamonds) -- to the middle point of the C-C bond.
} \label{fig2}
\end{figure}

We now consider the molecular complexes Th@C$_{20}$ and Th@C$_{20}$H$_{20}$.
The C$_{20}$ molecule with a dodecahedral cage structure is the
smallest member of the fullerene family \cite{C20}.
Experimentally, it has been found in the form of the dodecahedrane C$_{20}$H$_{20}$,
whose high symmetry shape is stabilized by the hydrogen passivation \cite{Paquette}.
Earlier \cite{Singh}, it has been shown that the thorium encapsulation stabilises the molecule Si$_{20}$
which is the silicon analogue of C$_{20}$.
Without the thorium atom the radius of the carbon cage is 2.04~{\AA} in C$_{20}$
and 2.17~{\AA} in C$_{20}$H$_{20}$ with the 1.09~{\AA} C--H bond length.
When Th is encapsulated in C$_{20}$ our calculations show that the radius
is increased to 2.23~{\AA}.
Note that this value is slightly smaller than the
characteristic distance (2.31~{\AA}) found for the optimal position of Th from
the C$_{60}$ fullerene sphere.
This indicates that in the C$_{20}$ fullerene the thorium atom most likely
resides at the molecular center.
Our calculations of the binding energy $E_b = E(\textrm{Th@C}_{20}) - E(\textrm{Th}) - E(\textrm{C}_{20})$ for Th displaced from the center of C$_{20}$,
fully confirms this observation, Fig.\ \ref{fig2}.
%
%
Note that unlike the displacement of Th inside C$_{60}$, Fig.\ \ref{fig1},
the situation with the position of Th in C$_{20}$, Fig.\ \ref{fig2}, is virtually isotropic
showing little difference for displacements in various directions.
The calculated energy minimum is 2.48 eV. Since $E_b > 0$ the calculated energy minimum
is metastable.

Our calculations for Th@C$_{60}$ and Th@C$_{20}$ indicate that the
situation with the optimal location of Th inside fullerenes is
somewhat similar to the behavior of C$_{60}$ in carbon nanotubes \cite{Mich}.
That is, the confinement of the thorium atom encapsulated in a fullerene
depends on its radius. In small fullerenes with radius $R < R_0 \approx 2.31$~{\AA}
the thorium atom is located at the molecular center as occurs in Th@C$_{20}$.
In fullerenes with a larger radius $R$ the thorium atom is displaced off
the molecular center staying at a distance of $R - R_0$ from it facing a hexagon
center, as occurs in Th@C$_{60}$. This conclusion is confirmed by our calculation
of Th@C$_{36}$ where the energy minimum is found for Th at a distance of 0.24 {\AA}
off the center while the estimated average radius of C$_{36}$ is 2.54~{\AA}.

Finally, the carbon cage radius in Th@C$_{20}$H$_{20}$ is 2.35~{\AA}, that is, larger than
the one in Th@C$_{20}$ whereas the local energy minimum of Th@C$_{20}$H$_{20}$
is increased to 8.14 eV. These data indicate that the complex Th@C$_{20}$H$_{20}$
is even less stable than Th@C$_{20}$.

\subsection{Th bonded to graphene (modelled by benzene and coronene)}
\label{Th-graph}

\begin{figure}
\resizebox{0.44\textwidth}{!} {
\includegraphics{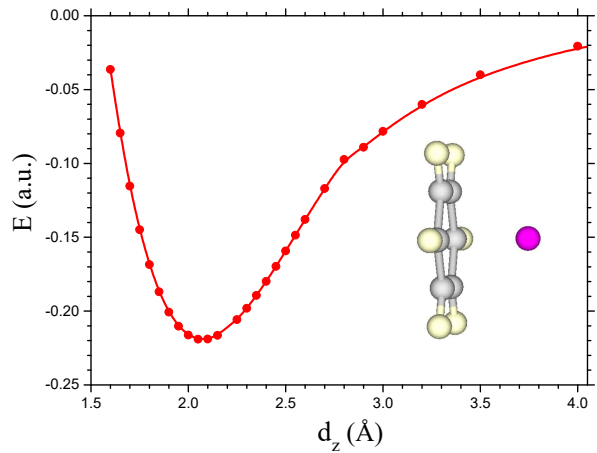}
}
\caption{
Binding energy $E_b$ of the Th--benzene molecular complex versus the Th displacement $d_z$ from the benzene (hexagon) center
in the perpendicular direction. The energy minimum ($-0.219$ Hartree or $-5.97$~eV) corresponds to $d_z^{min}=2.05$~{\AA}.
} \label{fig3}
\end{figure}

\begin{figure}
\resizebox{0.46\textwidth}{!} {
\includegraphics{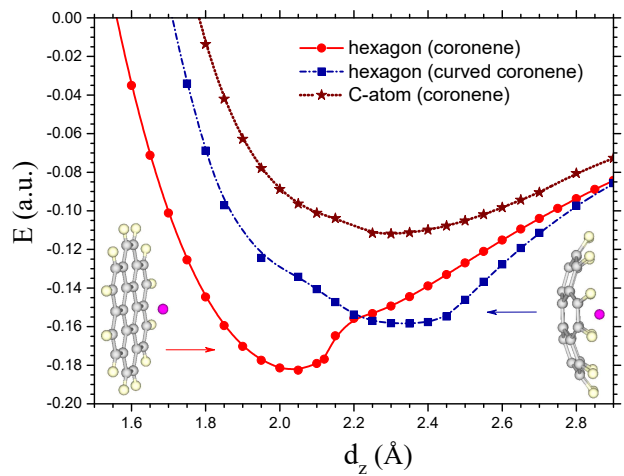}
}
\caption{
Binding energy $E_b$ of the Th--coronene molecular complex versus the Th displacement $d_z$ from the inner hexagon
in the perpendicular direction:
1) red line (circles) -- from the center of the hexagon ($d_z^{min}=2.02$~{\AA}),
2) blue line (squares) -- from the hexagon center of the curved coronene ($d_z^{min}=2.33$~{\AA}),
3) brown line (stars) -- from the C atom, see text for details.} \label{fig4}
\end{figure}

We further investigate the properties of the energy minimum found with the Th atom
facing the center of the hexagon in the fullerene C$_{60}$.
The most important question is how general is this chemical bonding revealed in Th@C$_{60}$.
For that purpose we consider the formation of chemical bonding in other molecular systems:
graphene and a carbon nanotube.
However, both graphene and a carbon nanotube are infinite objects, and below we model
them with simpler planar molecules like benzene (C$_{6}$H$_{6}$, insert of Fig.\ \ref{fig3}) and coronene.
Coronene (C$_{24}$H$_{12}$, the insert panel on the left in Fig.\ \ref{fig4}) is a polycyclic aromatic hydrocarbon
comprising seven carbon rings.
Note that the C-C bond length of the inner six-member carbon ring in coronene
is very close to the bond length (1.42 {\AA}) observed in graphene.
Interestingly, there are two-dimensional boron-nitride materials like
BN-circumacenes \cite{Mocci} isoelectronic with carbon-based planar molecular structures,
which can possibly accommodate thorium as well.
In Ref.\ \cite{Du17} the interaction of Th with graphene-like molecules has been
calculated for high spin (triplet and quintet) ground states of thorium.
Our study show that the singlet spin-adapted state is certainly preferable.
It results in smaller Th-C bond length: 2.48 {\AA} compared with 2.62 {\AA} for the triplet and 2.61 {\AA} for the quintet state, and a lower total energy.
The triplet-singlet energy difference is 1.623 eV, the quintet-singlet one is 2.025~eV (with the jorge-ADZP basis set \cite{BSE}).
The singlet ground state of Th has been also confirmed in Th@C$_{74}$ \cite{Li-b}.

\begin{figure}
\resizebox{0.46\textwidth}{!} {
\includegraphics{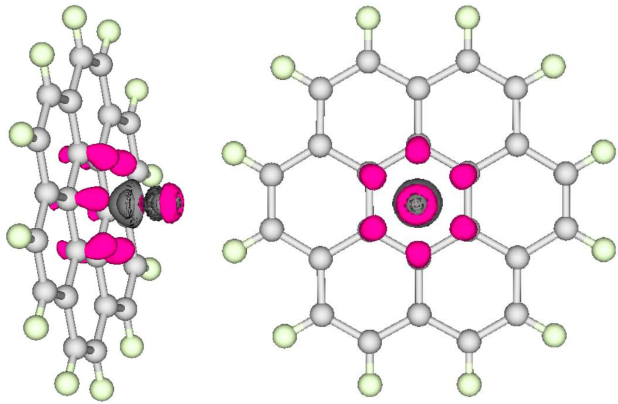}
}
\caption{
The chemical bonding of Th with coronene (C$_{24}$H$_{12}$) shown as
the bonding electron density $\rho_b = \rho(\mbox{Th+C}_{24}\mbox{H}_{12})-\rho(\mbox{Th})-\rho(\mbox{H}_{12}).$
Red clouds indicate an excess of the bonding electron density, black ones show a depletion of the electron density.
There is a partial charge transfer from thorium to coronene.
}
\label{fig5}
\end{figure}

The results of our singlet ground state calculations for Th bonded to benzene and coronene are shown
in Table \ref{tab1} and Figs.\ \ref{fig3}, \ref{fig4},~\ref{fig5}.
%
%
%
%
In both cases the geometry optimization without a symmetry restriction yields a well developed energy minimum
indicating a formation of chemical bonding
with the thorium atom located above the molecule center. We conclude that the molecular symmetry is $C_{6v}$
and adopt it for further considerations.
The closest Th-C bonds are 2.50 {\AA} (benezene) and 2.47 {\AA} (plain coronene).
Displacing Th in other directions - for example, above the C atom (brown line in Fig.\ \ref{fig4}) - also leads to a minimum, but in that case its energy is at least 1.94~eV higher that for the global one.
The chemical bonding developed between Th and the coronene molecule is visualized in Fig.\ \ref{fig5}.
%
%
In terms of molecular orbitals the situation can be interpreted similar to that for Th@C$_{60}$.
In benzene the LUMO state (at 3.97 eV for C$_6$H$_6$) belongs to the two-dimensional $E_2$ irreducible
representation of the $C_{6v}$ symmetry group.
In the presence of thorium the $E_2$ orbital being hybridized with the $Y_{l=2}^{m=\pm2}$ $d-$states and $Y_{l=3}^{m=\pm2}$ $f-$states
of thorium, is shifted to the bonding energy of -4.53 eV, Table \ref{tab1}, and becomes fully
populated by four Th valence electrons.

It is instructive to compare the bonding of Th with other four-valent metals: Ti, Zr and Hf, Table \ref{tab2}.
There is a remarkable difference between the bond lengths of Ti and Zr with C$_{6}$H$_{6}$ on the one hand,
and Hf and Th with C$_{6}$H$_{6}$ on the other. Out of the three metals, only hafnium
demonstrates the bond lengths close to that of thorium. It is also worth noting that
the difference is reflected in the character of the HOMO level. In HfC$_{6}$H$_{6}$ HOMO
is the two-fold $E_2$ molecular orbital like in ThC$_{6}$H$_{6}$ whereas in
TiC$_{6}$H$_{6}$ and ZrC$_{6}$H$_{6}$ it is the non-degenerate $A_1$ level.
%
\begin{table}
\caption{Comparison with other four-valent transition metals ($M$): Ti, Zr and Hf,
located above the benzene C$_6$H$_6$, MP2 calculations.
$d_0$ is the distance to the carbon ring ({\AA}),
 $d(M-$C) is the metal-carbon bond length ({\AA}), MO energies are in eV.
 (The metal basis sets are jorge-ADZP \cite{BSE}.)
\label{tab2} }

\begin{ruledtabular}
\begin{tabular}{c | c  c  c }

metal           & Ti     & Zr    & Hf    \\
\tableline
    $d_0$       & 3.232          & 3.386          &  1.809 \\
    $d(M-$C)    & 3.522          & 3.664          &  2.256 \\
    HOMO $-$ 1  & -8.923 ($A_1$) & -6.272 ($A_1$) & -9.793 ($E_1$) \\
     HOMO       & -5.178 ($A_1$) & -4.882 ($A_1$) & -4.903 ($E_2$) \\
     LUMO       & 1.720 ($E_1$)  & 0.985 ($A_1$)  & -0.261 ($A_1$) \\
    LUMO $+$ 1  & 2.626 ($E_2$)  & 1.347 ($E_1$)  & 1.001 ($A_1$) \\
\end{tabular}
\end{ruledtabular}
\end{table}

In addition, we have studied the behavior of the Th-C chemical bonding when the carbon network is curved.
The effect of curvature was modelled by placing the carbon and hydrogen atoms of coronene on
the spherical surface with the radius of 7.1 {\AA} which is twice the value of the radius of the C$_{60}$ fullerene.
The influence of curvature on the bonding is shown by the blue plot of Fig.\ \ref{fig4}.
We observe that the position of the energy minimum is shifted from 2.05 {\AA} to 2.33 {\AA},
whereas its value becomes $\approx$0.63 eV higher.
This rise of the energy for the curved coronene is accounted for by an increase of the electrostatic repulsion
between the positively charged thorium atom and terminated hydrogen atoms of coronene.
In real carbon systems like graphene or a carbon nanotube there are no positively charged hydrogen atoms
and we expect that this effect will be less pronounced.

\section{Conclusions}
\label{sec:con}

We have studied the optimal position of the Th atom with novel carbon materials
by means of the Hartree-Fock approach with weak dispersion interactions accounted for by the MP2 correction,
Table \ref{tab1}.
Having started with the Th atom
encapsulated in the fullerenes C$_{60}$, we find
the favourable location in the C$_{60}$ fullerene.
In this energy minimum state the Th atom faces the center of
the 6 member carbon ring and is at the distance of 2.01 {\AA} from the center.
The distances to first and second neighboring carbon atoms of C$_{60}$ in that case
are 2.46 and 2.85~{\AA}, respectively.

This finding is in agreement with results of DFT calculations of Th encapsulated in
fullerenes of various shape: for Th@C$_{84}$ \cite{Kamin}, Th@C$_{86}$ \cite{Wang19}
and Th@C$_{76}$ \cite{Jin19}. The closest Th-C bond lengths in these molecules
vary from 2.39 to 2.53 {\AA}.
In the C$_{20}$ and C$_{20}$H$_{20}$ metastable complexes the fullerene radius is too small and the Th atom
occupies the position at its center causing a small expansion of the outer carbon cage.

In C$_{60}$ we have also studied other local minima of the Th atom and
found that they are separated from the global hexagon minimum
by an energy gap of 0.22 eV.
Therefore, the global minimum is quite pronounced and the situation is
different from the very shallow energy minimum in La$_2$@C$_{80}$-$I_h$(7) \cite{popov}.

Finally, we have thoroughly investigated the nature of the global energy minimum for the Th atom bound to
a carbon network.
In that case the carbon cage was taken in the molecular form of benzene and coronene.
Th with benzene and Th with coronene give reasonable approximations to such molecular systems
as Th bound to graphene and Th bound to a carbon nanotube.
In benzene and coronene we have found well pronounced energy minima for Th located at distances of 2.05 and 2.02 {\AA}
from the center of the carbon hexagon, correspondingly.
(The triplet and quintet spin states of Th lying much higher in energy in comparison with the singlet-adapted ground state,
can be omitted from consideration.)
The shortest Th-C bond lengths are 2.50~{\AA} (benzene) and 2.47~{\AA} (coronene).
Moreover, in the case of curved coronene modelling a piece of a fullerene or a carbon nanotube, this global minimum survives but
the optimal distance from Th to the center of the hexagon ($d_z^{min}$) becomes larger.
In the graphene network the global minimum is deeper than for fullerenes, and the energy difference with other local minima is at least 1.7~eV.

\acknowledgements

This research was supported by a grant of the Russian Science
Foundation (Project No 19-72-30014).\\
Computations have been partially carried out using the
equipment of the shared research facilities of HPC computing
resources at Lomonosov Moscow State University
\cite{Sadovnichy-13}.


\newpage
\noindent {\bf REFERENCES}

\end{document}